\documentclass{elsarticle}

\usepackage{booktabs} 
\usepackage{amsmath}
\usepackage{amssymb}
\usepackage{tikz}
\usepackage{graphicx}
\usepackage[utf8]{inputenc}
\usepackage{algorithm}
\usepackage[noend]{algpseudocode}
\usepackage{stmaryrd} 

\usepackage{amsfonts}
\usepackage{url}

\newcommand{\tuple}[1]{\ensuremath{\langle #1\rangle}}

\newcommand{\nico}[1]{\textcolor{purple}{#1}}

\begin{document}

\begin{frontmatter}

\title{A Subjective Interestingness measure for Business Intelligence explorations}

\cortext[mycorrespondingauthor]{Corresponding author}

\author[ut]{Alexandre Chanson}
\ead{alexandre.chanson@etu.univ-tours.fr}

\author[ut]{Ben Crulis}
\ead{ben.crulis@etu.univ-tours.fr}

\author[ut]{Nicolas Labroche}
\ead{nicolas.labroche@univ-tours.fr}

\author[ut]{Patrick Marcel\corref{mycorrespondingauthor}}
\ead{patrick.marcel@univ-tours.fr}

\address[ut]{University of Tours, Tours, France}

\begin{abstract}
This paper addresses the problem of defining a subjective interestingness measure for BI exploration. Such a measure involves prior modeling of the belief of the user. The complexity of this problem lies in the impossibility to ask the user about the degree of belief in each element composing their knowledge prior to the writing of a query. To this aim, we propose to automatically infer this user belief based on the user's past interactions over a data cube, the cube schema and  other users' past activities.  We express the belief under the form of a probability distribution over all the query parts potentially accessible to the user,
and use a random walk to learn this distribution. This belief is then used to define a first Subjective Interestingness measure over multidimensional queries. Experiments conducted on simulated and real explorations show how this new subjective interestingness measure relates to prototypical and real 
user behaviors, and that query parts offer 
a reasonable  proxy to infer user belief.

\end{abstract}

\end{frontmatter}

\section{Introduction} \label{sec:intro}
Business intelligence (BI) exploration can be seen as an iterative process that involves expressing and executing queries over multidimensional data (or cubes) and analyzing their results,  to ask more focused queries to reach a state of knowledge that allows to answer a business question at hand. This complex task can become tedious, and for this reason, several approaches have been proposed to facilitate the exploration by
pre-fetching data \cite{DBLP:conf/dawak/Sapia00},
detecting interesting navigation paths \cite{DBLP:journals/vldb/Sarawagi01},
recommending appropriate queries based on past interactions \cite{DBLP:journals/dss/AligonGGMR15} or by modeling user intents \cite{DBLP:conf/caise/DrushkuALMPD17}. 

Ideally, such systems should be able to measure to which extent a query would be interesting for a given user prior to any exploration. Indeed, as illustrated in \cite{IDEA:2018} and first elicited in \cite{DBLP:conf/kdd/SilberschatzT95} in the context of Explorative Data Mining (EDM), the interestingness of a pattern
depends on the problem at hand, and, most importantly, on the user that extracts the pattern.
An interestingness measure for such explorative tasks should  therefore be
tailored for a specific user. 

Following the idea of subjective interestingness measures initiated and developed by De Bie 
\cite{DBLP:conf/ida/Bie13}, our aim is to measure
the subjective interestingness of a query a expressed by a coherent set of query parts, 
 based on the prior knowledge that the user has about the cube and the cost for the user to understand the query and its evaluation. 


It is therefore crucial, before reaching the definition of such an interestingness measure for BI, to be able to transcribe, with an appropriate information-theoretic formalism, the prior user knowledge, also called belief, on the data. 
De Bie proposes to represent this belief as a probability distribution over the set of data. However, 
it is clearly not possible to explicitly ask a user about the degree of belief in each element composing their knowledge prior to each query, let alone 
identifying on which element of knowledge expressing this probability distribution. This motivates the investigation of approaches for automatically estimating the user's belief
based on their implicit feedback.
Let us now consider the following example to illustrate the difficulty of estimating probabilities for the belief.

\begin{figure}[ht]
    \centering
    \includegraphics[scale=0.5]{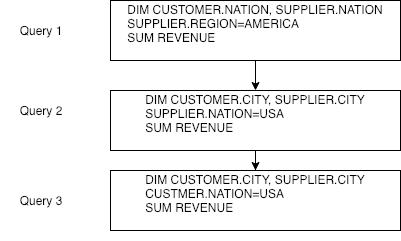}
    \caption{Toy SSB benchmark session}
    \label{fig:ssb}
\end{figure}

\paragraph{Example} Consider the  exploration  over the schema of the  Star Sche\-ma Benchmark  \cite{DBLP:conf/tpctc/ONeilOCR09}, consisting of $3$ queries, as illustrated in Figure \ref{fig:ssb}, and loosely inspired by session 3 of the SSB's workload. 
For the sake of readability, only the relevant query parts (grouping set, filters and measures) are shown.
This  example showcases the short session initiated by a user that explores the cube looking for information on revenue some company makes in different locations. 
Assume we are interested in recommending a query to the user
for continuing the exploration. This recommendation should be both
connected to the most used query parts, so as not to loose focus,
but also should bring new, possibly unexpected information, so as
not to feed the user with already known or obvious information.
A naive solution would be to use the set of all possible query parts as the set of data and to express the belief based on the frequency of each query part in the past user history. 
From the session in Figure \ref{fig:ssb} it is possible to compute the number of occurrences of each query part (for instance, SUM REVENUE appears 3 times, 
CUSTOMER.CITY 2 times, while SUPPLIER.REGION=AMERICA appears only once, etc.).
However, this simple representation raises major problems: first, the vector of user belief computed from the number of occurrences will mostly contain 
zero values because the majority of users will concentrate their exploration to a certain region of the data cube. Second, this belief would not give any probability to query parts such as CUSTOMER.NATION=CANADA, while if user knows about \textit{AMERICA} and \textit{USA}, she is likely to have a basic knowledge about sibling countries to \textit{USA} in the dimension \textit{CUSTOMER.NATION}. Finally, it may also be taken advantage
of  other users' former explorations, as a proxy of what the current user might find interesting.

This example stresses the need for an approach to define the belief based on the users' past activity, as well as an information about how knowledge is structured, which, in the case of the data cube, can be found in the cube schema.
We note that while previous works already investigated surprising data in cubes (see e.g., 
\cite{DBLP:journals/vldb/Sarawagi01,DBLP:journals/ijbidm/CariouCDGGK09}), to the best of our knowledge none of them did so by explicitly modeling a user's belief.

As a first step in this direction, our previous paper \cite{DBLP:conf/dolap/ChansonCDLM19} tracks user belief in BI interactions for measuring the subjective interestingness of a set of queries executed on a data cube. This approach builds a model of the user's past explorations that is then used to infer the  belief of the user about the query parts being useful for the exploration. 
Contrary to the context of pattern mining 
\cite{DBLP:conf/ida/Bie13} where in general no metadata  information is available, the query parts that are employed in this model cannot be considered agnostically of the cube schema,
that the user typically knows. In this context, the method introduced in 
 \cite{DBLP:conf/dolap/ChansonCDLM19}
 takes advantage of the schema to infer what a user may or may not know based on what has been already visited and what is accessible from the previous queries.
The  belief of a user is defined as a probability distribution over the set of query parts coming from the log of past activities and the cube schema. This probability distribution is learned as the resulting stationary distribution of a modified topic-specific PageRank algorithm, where the underlying graph topology matrix is based on previous usage and the schema of the cube, and where a teleportation matrix that corresponds to a specific user model is introduced to 
ensure connectivity.
Finally, \cite{DBLP:conf/dolap/ChansonCDLM19} takes advantage of the artificial exploration generator CubeLoad \cite{DBLP:conf/caise/RizziG14} that mimics several prototypical user behaviors to evaluate qualitatively and quantitatively divergences in the estimated user belief.
%
%

The work presented here improves on that introduced in \cite{DBLP:conf/dolap/ChansonCDLM19} with several major contributions, both at the methodological level and at the experimental level: 
\begin{itemize}
    \item it refines the belief model by taking into account the filter values of the query parts. This is a strong bottleneck since it adds a lot of vertices in the graph used to compute user belief. To solve this problem, new rules to build the graph are introduced in Section \ref{sec:PRbelief} ;
    
    \item as stated before, this new belief model benefits from information from the cube schema, the past logs and an actual user profile  represented by her explorations. All the underlying operations are now defined as simple graph manipulation primitives and allow for an easier understanding of the whole process ; reproducibility of the experiments 
    is permitted thanks to the shared code repository\footnote{\url{https://github.com/AlexChanson/IM-OLAP-Sessions}} ;
    
    \item according to these new construction rules, the approach now results in only one strongly connected graph which avoids the need for 
    a teleportation matrix and ensures a cleaner convergence mechanism. Details of the simplified PageRank algorithm is provided in Section \ref{sec:pr_section} ; 
    
    \item as in \cite{DBLP:conf/dolap/ChansonCDLM19}, we introduce a real valued parameter $\alpha \in [0,1]$ that allows to give more or less importance to the user specific exploration bias in the computation of the belief distribution. This parameter is at the core of our experiments as it allows to reveal, when set close to $1$, significant differences in belief and subjective interestingness models ;
    
    
    \item our approach now contains a simple yet efficient incremental mechanism that allows to track the evolution of the belief during an exploration  as described in Section \ref{sec:inc_belief} ; 
    
    \item finally, this paper introduces a first formalization of a subjective interestingness measure for Business Intelligence explorations based on the belief distribution and on a simple measure of the complexity of  a query formed by several query parts, as described in Section \ref{sec:subj_interest_dopan}.
\end{itemize}

As in \cite{DBLP:conf/dolap/ChansonCDLM19}, one difficulty stands in the evaluation of our proposal, as there is no ground truth available. In this context, we propose several experiments:
\begin{itemize}
    \item an updated qualitative and a quantitative evaluation of the belief distributions for several simulated user profiles using CubeLoad generator \cite{DBLP:conf/caise/RizziG14} and  new experiments on real data with the DOPAN workload \cite{DjedainiDLMPV19} ;
    \item a set of novel experiments that a posteriori estimates the subjective interestingness of queries in simulated and real explorations from CubeLoad and the DOPAN workload. 
\end{itemize}

Experimental conclusions show that our approach to model user belief and subjective interestingness from a graph of query parts: ($1$) behaves as expected on both prototypical 
user behaviors and real user explorations, 
and ($2$) indicates that query parts are a good proxy to infer user belief.


This paper is organized as follows: Section \ref{sec:example} motivates 
the use of user belief and subjective interestingness measure in the context of 
BI exploration.
Section \ref{sec:preliminaries} introduces the concepts used in our approach:  formal definitions  BI explorations, query parts and concepts related to Subjective Interestingness and PageRank algorithm. Section \ref{sec:IM} introduces the graph based user belief model using past explorations and schema as inputs, and introduces a novel algorithm to deal with incremental belief estimation. Section \ref{sec:subj_interest_dopan} introduces the Subjective Interestingness definition based on said model. Finally, Sections \ref{sec:tests} and \ref{sec:test_si_dopan} present our experiments to assess  our belief model and our Subjective Interestingness measure, both on artificially generated  explorations and on real user explorations. Finally, Section \ref{sec:related} discusses related work and 
Section \ref{sec:conclusion} concludes and draws perspectives.

\section{Our vision of User Centric Data Exploration}
\label{sec:example}

This section describes how the knowledge of a user belief, and by extension a subjective interestingness measure, could be used to improve the user's experience in the context of interactive data exploration. 
This example highlights the main scientific challenges of such task, some of them being left as future work as the present paper exclusively focuses on a first expression of user belief and a derived subjective interestingness measure in the context of data cube exploration.

\begin{figure*}[ht]
    \centering
    \includegraphics[scale=0.5]{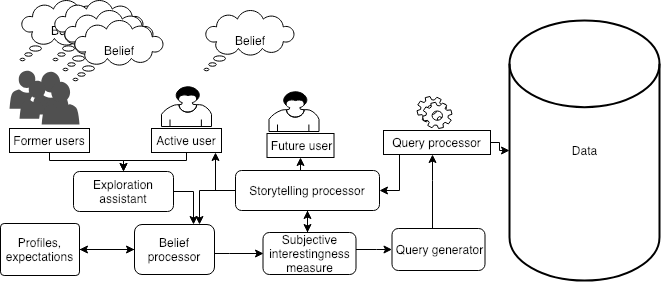}
    \caption{Envisioned use of belief and subjective interestingness measures in data exploration
    }
    \label{fig:reco_interestingness}
\end{figure*}

In our vision, illustrated in Figure \ref{fig:reco_interestingness},
human remains in the loop of data exploration,
i.e., the exploration is not done fully automatically,
but we aim at making it less tedious.
All users, naive or expert,
willing to explore a dataset,
express their information need 
through an exploration assistant.
This assistant is left with the task
of deriving from the user's need the
actual queries to evaluate over the
data source.
This exploration assistant communicates
with a belief processor that 
is responsible for the maintenance
of the user's profile, i.e., a model
of that user, in the sense that 
it includes an estimation  of the actual 
belief unexpressed by the user.
This belief is manifold and concerns e.g., 
hypotheses on the value of the data,
the filters to use, how the answer should be
presented, etc.
The belief processor  activates a series of 
subjective interestingness measures that drives
the query generator for deriving 
and recommending the most
interesting queries for this user, in the sense that
they produce relevant, unexpected, diverse answers, avoiding 
undesirable artifacts such as  
biased or false discoveries, 
the so-called cognitive bubble trap, etc.
These answers and recommendations are packaged 
(e.g., re-ranked, graphically represented)
by the storytelling
processor before being displayed to the user and
sent to the belief processor for profile updating.

Notably, thanks to the belief processor,
once enough diverse users are modeled, 
the storytelling processor may 
cope with the  cold start problem of 
generating recommendation
for unknown users 
(the future user of Figure \ref{fig:reco_interestingness}),
e.g., by removing bias introduced by
common beliefs.

The work presented in this paper is the first step in the implementation of this vision. We first concentrate on cube exploration, expressing the belief over query parts and deriving an incremental Subjective Interestingness measure on queries. Noticeably, all our definitions take advantage of the peculiarities of the data cube exploration context to be on par with what a human analyst would consider interesting.

\section{Preliminaries}\label{sec:preliminaries}

This section introduces the basic definitions of our framework.

\subsection{BI explorations}

Our work considers BI explorations,
i.e., sequences of OLAP queries over a 
database instance under a star schema, called datacube.

Let $D$ be a database schema, $I$ an instance of $D$ and $Q$ the set of formal queries one can express over $D$. For simplicity, in this paper, we consider relational databases under star schemata, queried with multidimensional queries.
Let $A$ be the set of attributes of the relations of $D$. Let $M \subset A$ be a set of attributes defined on numerical domains called measures.
Let 
$H = \{h_1 , \ldots, h_n \}$ be a finite set of hierarchies, each characterized by (1) a subset $Lev(h_i)$ of attributes called levels and (2) a roll-up total order $\succeq_{h_i}$ of $Lev(h_i)$. 
We denote by $adom(A)$ the set of all constants appearing the instance $I$ of $D$ for attribute $A$.
For each hierarchy $h_i$, $Lev(h_i)$ includes
a top-most level $l_i^{ALL}$ such that $\nexists l\in Lev(h_i), l_i^{ALL} \succeq_{h_i} l$. 
This level only has one value called $all_i$, i.e., $adom(l_i^{ALL})=\{all_i\}$.
For any two consecutive levels $l^1_i, l^2_i$ of a hierarchy $h_i$,
function $children(m)$ applied to $m\in adom(l^1_i)$ returns
the set of values in $adom(l^2_i)$
that are direct children of $m$ according to $h_i$.

To simplify, we describe an OLAP query $q$ in $Q$ as a set of query parts.  Note that the term query parts can undergo different meanings. Coherent with our objective of taking into account both usage (i.e., previous explorations) and cube schema, our query part definitions encompasses both.
We rely on the definition of query part provided by \cite{DBLP:conf/caise/RizziG14}, where the authors consider it is one
constituent of a 
\begin{quote}
multidimensional query consisting of (i) a group-by (i.e., a set of hierarchy levels on which measure values are grouped); (ii) one or more measures whose values are returned (the aggregation operator used for each measure is defined by the multidimensional schema); and (iii) zero or more selection predicates, each operating on a hierarchy level.    
\end{quote}

However, in our case, a query part is not necessarily  attached to a query already expressed by some user,  since we aim at considering also query parts that might be used in the future.

Formally, a query part is either (i) a level $l$ of a hierarchy in $H$, (ii) a  measure in $M$, or the member $v$ of a simple Boolean predicate of the form  $l = v$, where $l$ is a level of a hierarchy in $H$, and $v$ is a constant in $adom(l)$. 
Note that each member $v$ identifies its level and hierarchy.
Given a database $D$  we call $P_D$ the set of query parts.
In what follows, queries are confounded with their sets of query parts, unless otherwise stated, and we assume a function $parts(q)$ that applied over a query $q$ returns the subset of $P_D$ 
containing its query parts.



Finally, a BI exploration $s$  is a sequence $[q_1, \ldots, q_p]$  of $p$ OLAP queries, and a log is a set of explorations.




\subsection{Interestingness for exploratory data mining} \label{sec:DeBie}

The framework proposed by De Bie \cite{DBLP:conf/ida/Bie13}, in the context of exploratory data mining, is based on the idea that the goal of any exploratory data mining task is to pick patterns that will result in the best updates of the user's knowledge or belief state, while presenting a minimal strain on the user's resources. 
In De Bie's proposal, the belief is defined for each possible value for the data from the data space and can be approximated by a background distribution.

As a consequence, a general definition for this interestingness measure (IM) is a real-valued function of a background distribution, that represents the belief of a user, and a pattern, that is to say the artifact to be presented to the explorer. Given a set $\Omega$, the data space, and a pattern $\Omega'$ a subset of $\Omega$, the belief is the probability P$(\Omega')$ of the event  $x \in \Omega'$, i.e., the degree of belief the user attaches to a pattern characterized by $\Omega'$ being present in the data $x$. In other words, if this probability is small, then the pattern is subjectively surprising for the explorer and thus interesting. In this sense, the IM is subjective in that it depends on the belief of the explorer. De Bie also proposes to weight this surprise by the complexity of the pattern $\Omega'$ as follows:
\begin{equation}\label{eq:IM_DeBie}
    IM_{De Bie}(P,\Omega') = \frac{-log(P(\Omega'))}{descComp(\Omega')}
\end{equation}
where $P$ represents the user belief, i.e., the background distribution of the pattern $\Omega'$ over the set of data $x$ and $descComp(\Omega')$ denotes the description complexity of a pattern $\Omega'$.

The data mining process consists in extracting patterns and presenting first those that are subjectively interesting, and then refining the belief background distribution based the newly observed pattern $\Omega'$.
The key to such modeling as proposed by De Bie lies in the definition  of the belief of each user for all possible patterns and how it should evolve based on new patterns explored during time. 

\subsection{PageRank}
\label{sec:pr_section}
Initially, the PageRank algorithm is designed to estimate the relative importance of web pages as a probability to end up on this web page after an infinite surf on the web \cite{DBLP:journals/cn/BrinP12}. This algorithm is based on ergodic Markov Chains. A Markov Chain models a collection of states and transitions probabilities from a state to the next. At a particular time t, a random walker is assumed to be in a unique state of the chain. We are interested by the probability distribution over the states after some time. After an infinite amount of time, this distribution  is called the stationary distribution, if it exists. As the transitions probabilities themselves cannot change with time, these chains can be represented as simple directed graphs with the different possible states as nodes and the transitions as edges with their assigned probabilities. A Markov Chain can also be represented as an (N,N) matrix, where N is the number of possible states. A Markov Chain that is aperiodic and where all states are connected with all other states by a sequence of states whose transitional probabilities are not zero has the ergodic property. In other words, it is possible to reach any state from any initial state with enough time. This implies that a stationary distribution over the possible states exists and is unique after an infinite number of iterations, independently of the starting state.

In the classical Page Rank algorithm used for the web, the pages are the possible states of the Markov Chain and the hyperlinks are the transitions between states.
The transitional probabilities from a particular page are proportional to the number of hyperlinks targeting a common page. Thus, to rank the pages by popularity, one needs to find the stationary distribution over the pages. These are the probabilities of landing on a page after surfing for a long time, starting from any page.
The stationary distribution, called the PageRank vector $PR$, is the solution to the following equation:
\begin{equation}
\label{eq:PR}
PR =  M \times PR 
\end{equation}
where $M$ is the stochastic transition matrix of the graph of web pages hyperlinks. 


Our approach considers query parts as
states in a Markov chain and
Section \ref{sec:PRbelief} explains how
transitional probabilities are defined.


\section{Inferring user belief from schema and log usage}
\label{sec:IM}

This section presents our first contribution and addresses the following questions: ($1$) what is user belief in BI exploration? ($2$) How to estimate it? ($3$) How to make it evolve during an exploration?

\subsection{What is user belief in BI?}\label{sec:beliefBI}

Ideally, in the context of BI exploration, the user belief would be a probability the user attaches to the statement "I believe the value of this cell is exactly this one". Modeling such a belief is one of our long term perspectives, as such this would raise several questions such as: ($1$) how to ensure efficiency while executing all the queries to update the belief over cells? ($2$) How to cope with the combinatorial complexity of expressing the belief over subsets of cells rather than a single cell? ($3$) How to update the belief distribution over cells when observing  aggregated values at higher level of granularity? 
As an example, Sarawagi in her seminal work \cite{DBLP:conf/vldb/Sarawagi00} 
to estimate user belief,
restricts her study to the sum measure paired with an assumption of uniform distribution to locally estimate belief on cells when exploring aggregates.
Sarawagi's work relies on the following assumptions: belief is only expressed
over a limited set of cells (relating to those already explored) and the
cube instance and past query answers are available to estimate this belief.

\begin{figure}[ht]
    \centering
    \includegraphics[scale=0.3]{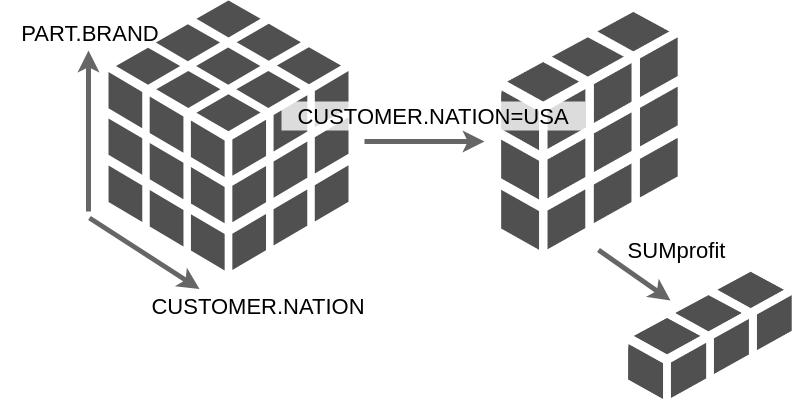}
    \caption{Aligned with De Bie's framework, query parts can be seen as restrictions to the original data space in the case of an OLAP cube exploration}
    \label{fig:querypart_asrestriction}
\end{figure}

In a first methodological step towards this ambitious direction, 
we use query parts as proxies to estimate this user belief.
We consider in this work that the user belief is the importance  the user attaches to the statement "I believe this query part is relevant for my exploration". 
In some sense, we consider query parts as
pieces of knowledge about the data that
reduce the set of possible values it may take from the original data space, inspired by the De Bie's view of explorative pattern mining 
\cite{DBLP:conf/kdd/Bie11,DBLP:journals/ml/KontonasiosB15} and as illustrated in Figure \ref{fig:querypart_asrestriction}.

We propose to define the user belief over the set of query parts for the following reasons. First, the set of query parts is measurable (and usually reasonable in size) and thus respects the formal constraints in the model of De Bie \cite{DBLP:conf/ida/Bie13} to extend the belief to an interestingness measure. 
Second, database instance or query answers may not be available,
e.g., for privacy or confidentiality reasons, when query logs are anonymized. Finally, query parts provide a finer level to work with compared to queries.
Working at the query level would end-up with a very sparse representation of the data space, as the probability that two queries occur in the same exploration is much lower than the probability that two query parts appear in the same query or exploration. 
Moreover, when considering query parts, the most interesting ones for the user may appear in several consecutive queries and thus might have more prominent probability values.

As we cannot "brain-dump" the user, the belief is approximated by the importance of the available query parts. The challenge lies in a way to find this probability distribution over a possibly infinite or too large set of query parts even if we restrict to the attributes in a given schema.
Practically, in order to avoid to deal with all these query parts, we restrict to those appearing  in a query log or in the schema, where only the active domain of the attributes is considered. 
Subjectiveness is ensured by the importance attached to the query parts appearing of the user's log of former explorations.

\subsection{Using PageRank as a belief function}
\label{sec:PRbelief}
Once restricted the set of query parts
, we still need to compute their relative importance expressed as a probability distribution for a specific user. As explained previously, this is done by a PageRank (PR) algorithm that computes the probability 
for a user $u$ to end up on a query part $p$ when using the cube schema during the exploration, knowing past explorations by other users and knowing the profile of $u$.
A naive assumption that could be made on this set of parts for an initial background distribution is that all parts not seen by the user are equally probable and those seen are as probable as often they appear in the user's log. This  would ignore many behaviors evidenced in the user explorations and connections of parts in  the schema of the cube. Our approach  incorporates those elements.


Given a database schema $D$ with query parts $P_D$,
the input to the PageRank algorithm is 
a directed graph of query parts $G = \tuple{P_D,E}$,
 computed by Algorithms \ref{algo:graph_building_schema}, 
 \ref{algo:graph_building} 
and  \ref{algo:eval}
respectively, as detailed in the following paragraphs.


Note that compared to our preliminary work \cite{DBLP:conf/dolap/ChansonCDLM19}, 
we use a more elaborated technique for building the graph.
In particular, we now consider as vertices the filter values (i.e., members) of potential selection predicates instead of the hierarchy levels on which they apply.
This brings richer information about the data in our approach without changing the overall method described in \cite{DBLP:conf/dolap/ChansonCDLM19}. The relationships between the selection predicates and their associated levels in the hierarchy are conserved but transcribed into the edges (see below and in Algorithm \ref{algo:graph_building}).


\begin{algorithm}
\caption{Schema based graph construction}\label{algo:graph_building_schema}
\begin{algorithmic}[1]
\Function{BuildSchemaGraph}{$D$}\\
\algorithmicrequire  A schema $D$ \\
\algorithmicensure  A graph of query parts
\State $V \gets P_D$, $E \gets \emptyset$
    \For{$h_i \in H$} \Comment{For each hierarchy in the schema}
            \State $E \gets LinkMember(E, \{all_i\})$ \Comment{Connects members}
        \For{$m \in adom(l)$ such that $l \in Lev(h_i)$}
                \State $E \gets E \cup \{\tuple{m,l,1}, \tuple{l,m,1} \}$
            \EndFor
        \For{$l^1_i,l^2_i \in Lev(h_i)$ such that $l^1_i \succeq_i l^2_i$ and $\nexists l^3_i, l^1_i \succeq_i l^3_i \succeq_i l^2_i$}
              
            \State $E \gets E \cup \{\tuple{l^1_i, l^2_i, 1}, \tuple{l^2_i, l^1_i, 1}\}$
            
        \EndFor
    \EndFor
\State \Return $\langle V,E \rangle$
\EndFunction

\Function{LinkMember}{E, m} \Comment{Recursively scans the hierarchy tree}
    \State $C \gets children(m)$
    \If{$C = \emptyset$}
        \Return E
    \EndIf
    \For{$c \in C$}
        \State $E \gets E \cup \{\tuple{m, c, 1}, \tuple{c, m, 1}\}$
        \State $E \gets LinkMember(E, c)$ 
    \EndFor
    \Return E
\EndFunction

\end{algorithmic}
\end{algorithm}

\begin{algorithm}
\caption{Log based Graph construction}\label{algo:graph_building}
\begin{algorithmic}[1]

\Function{BuildLogGraph}{$L,G$}\\
\algorithmicrequire  A log $L$ and a graph $G=\tuple{V,E}$ \\
\algorithmicensure  A graph of query parts
\For{$s=[q_1, \ldots, q_p] \in L$}
    \For{$i \in \llbracket 1, p-1 \rrbracket $}
        \State $P_i \gets parts(q_i)$, $P_{i+1} \gets parts(q_{i+1})$
        \State $V \gets V \cup P_i \cup P_{i+1}$
        \For{$p_1 \in P_i$}
            \For{$p_2 \in P_i$}
                \If{$\tuple{p_1, p_2, v}$ in $E$}
                    \State $E \gets (E \setminus \{\tuple{p_1, p_2, v}\}) \cup \{\tuple{p_1, p_2, v+1}\}$
                \Else
                    \State $E \gets E \cup \{\tuple{p_1, p_2, 1}\}$
                \EndIf
            \EndFor
            \For{$p_{next} \in P_{i+1}$}
                \If{$\tuple{p_1, p_{next}, v}$ in $E$}
                    \State $E \gets (E \setminus \{\tuple{p_1, p_{next}, v}\}) \cup \{\tuple{p_1, p_{next}, v+1}\}$
                \Else
                    \State $E \gets E \cup \{\tuple{p_1, p_{next}, 1}\}$
                \EndIf
            \EndFor
        \EndFor
    \EndFor
\EndFor
\State \Return $\tuple{V,E}$
\EndFunction

\end{algorithmic}
\end{algorithm}

\paragraph{Schema based construction rules} 
To represent the global topology induced by a database schema $D$, a  graph is constructed as follows,:
(i) there is an edge between any two consecutive levels of a hierarchy ;
(ii) for any there is an edge between a member and its direct children in the hierarchy of this member ;
(iii) finally, there is an edge between each member  and its
level attribute.
Details about the implementation of these rules are provided in Algorithm \ref{algo:graph_building_schema}.

\paragraph{Log usage construction rules} 
To represent the activity of a user or group of users, a graph can be constructed as follows.
There is an edge $e$ from query part  $p_1$ to query part $p_2$ and an edge $e'$ from $p_2$ to $p_1$ if $p_1$ and $p_2$ appear together in the same query. There is also an edge $e$ from query part $p_1$ to query part $p_2$ if $p_1$ is in a query that precedes, in an exploration, another query where $p_2$ appears. As described in Algorithm \ref{algo:graph_building}, those rules 
 can be applied either to generate the graph of all users past queries or to produce the graphs of a specific user by restricting the log used as input. Note that  Algorithm \ref{algo:graph_building} can either be used to update a pre-existing schema graph or, if the input graph is set to an empty graph, only build a new graph related to usage.
 
 
\paragraph{Introducing subjectivity}
Algorithms \ref{algo:graph_building_schema} and \ref{algo:graph_building} can be used to construct a graph that represents a general topology of the query space 
(Algorithm \ref{algo:graph_building_schema})
and transcribes important relationships established by the past users explorations (Algorithm \ref{algo:graph_building}, called with a query log detailing the past activities of all users). This graph $G_t$,
called the {\em topology graph} from now on,
is  however not subjective in any way. It has to be biased toward a specific user $U$, represented by the subset  of the query parts occurring in their sessions.
To this end, Algorithm \ref{algo:graph_building} can be called
over the query log of user $U$, and  defines $G_u = \tuple{P_D, E_u}$ called the {\em specific subjective user centered graph}.



\paragraph{Constructing the PageRank graph}
Once we have a graph representing the topology induced by the schema and the past logs graph, $G_t$, and a specific subjective user centered graph, $G_u$, we can aggregate them to produce the graph that will serve as an input for the PageRank algorithm described in Section \ref{sec:pr_section}. To that aim, Algorithm \ref{algo:merge_graph} introduces a real parameter $\alpha \in [0,1]$ that  allows to give more or less weight to graph $G_u$ compared to the topology graph $G_t$. Indeed, the topology graph is generally very large and the subjective user centered graph only modifies a small portion of it which may be barely noticeable in terms of belief distribution. In that sense, $\alpha$ can be seen as a normalization factor, as it can be used to control the relative importance of the user's log against the general log and topology inherited from the schema. 

\renewcommand{\algorithmicrequire}{\textbf{Input:}}
\renewcommand{\algorithmicensure}{\textbf{Output:}}

\begin{algorithm}
\caption{Graphs merging algorithm}\label{algo:merge_graph}
\begin{algorithmic}
\Function{Merge}{$G_1, G_2, \alpha$}\\
\algorithmicrequire ~2 graphs $G_1 = \tuple{V_1, E_1}$ and $G_2 = \tuple{V_2, E_2}$ and a real value $\alpha \in [0,1]$\\
\algorithmicensure ~a merged graph $G = \tuple{V, E}$
    \State $V \gets V_1 \cup V_2$ \Comment{Initialize the set of vertices of new graph}
    \For{$\tuple{p_1, p_2, v} \in E_1$} \Comment{Add all updated edges from $G_1$}
        \State $E \gets E \cup \{\tuple{p_1, p_2, (1 - \alpha) \times v}\}$
    \EndFor
    \For{$\tuple{p_1, p_2, v_2} \in E_2$} \Comment{Update with edges from $G_2$}
        \If{$\tuple{p_1, p_2, v_3}$ in $E$} 
            \State $E \gets (E \setminus \{\tuple{p_1, p_2, v_3}\}) \cup \{\tuple{p_1, p_2, v_3 + \alpha \times v_2}\}$
        \Else
            \State $E \gets E \cup \{\tuple{p_1, p_2, \alpha \times v_2}\}$
        \EndIf
    \EndFor
    \State \Return $\langle V,E \rangle$
\EndFunction

\end{algorithmic}
\end{algorithm}






Finally, the probability distribution over the set of query parts $P_D$ is computed as the PageRank vector on the graph resulting from Algorithm \ref{algo:merge_graph}. 
This vector is obtained by an iterative approach that converges after a sufficient number of iterations.





\paragraph{About the connectivity of the graph} This  approach has the advantage to produce a graph with only one connected component under the weak assumption that the log contains at least one query for each measure (i.e. each measure must appear in the log at least once),
which is a direct consequence of graph construction by Algorithms \ref{algo:graph_building_schema} and \ref{algo:graph_building}.
This property is crucial since it allows to simplify the Topic Specific PageRank algorithm used in \cite{DBLP:conf/dolap/ChansonCDLM19} into a more conventional PageRank algorithm as introduced in Section \ref{sec:pr_section}. Moreover, the construction of the 
aggregated graph that represents the logs, the schema and a specific user allows to tune the system to give more weight to any of these aspects.

\subsection{Incrementality of belief}\label{sec:inc_belief}

Contrary to most implementations of De Bie's framework, we don't enumerate possible patterns to recommend to the user. These implementations are able to update the beliefs assuming that each previous pattern has been seen and understood by the user without any user interaction. Instead, we recompute the estimated user beliefs as the user makes new queries which in turn will allow to recompute the Subjective Interestingness as described in Section \ref{sec:subj_interest_dopan}. In our experiments, we develop an a-posteriori method which quantifies the subjective interest of queries at any point of the exploration according to the previous queries that we know were launched by the user.

Let $\tuple{s,u}$ be a session defined as a sequence $s$ of queries and the user $u$ that produced this session and let $G_s = \tuple{V_s, E_s}$ be the \textit{active session graph}. The latter is constructed 
with 
Algorithm \ref{algo:graph_building}, {\tt BuildLogGraph}, 
with parameter $L$ 
restricted to 
the queries executed 
at some point of 
the user session $s$. By applying  Algorithm \ref{algo:graph_building} iteratively each time a new query is issued in session $s$,  with input parameter $G_s$ and a log restricted to the new query,  an updated version of $G_s$ is obtained.




Then,  Algorithm \ref{algo:merge_graph}
is applied to aggregate
the updated graph  $G_s$  with the topology graph $G_t$ 
using the {\tt merge($G_t$, $G_s$, $\alpha$)} method 
and the value  of  $\alpha$ to control how much of the active session influences the computation of the PR.

Finally, executing the PR algorithm on this aggregated graph leads to the updated belief distribution. Algorithm \ref{algo:eval} shows how this incremental computation of belief is implemented to define the expected subjective interestingness measure.



\renewcommand{\algorithmicrequire}{\textbf{Input:}}
\renewcommand{\algorithmicensure}{\textbf{Output:}}

\begin{algorithm}
\caption{Evaluation algorithm}\label{algo:eval}
\algorithmicrequire A session $s$ to be evaluated (as an ordered list of queries), the global log $L$, a database schema $D$ and a real value $\alpha \in [0,1]$\\
\algorithmicensure The subjective interestingness of each query of $s$
\begin{algorithmic}[1]

\State $G_t \gets  BuildSchemaGraph(S)$ \Comment{see Algorithm \ref{algo:graph_building_schema}}
\State $G_t \gets  BuildLogGraph(L, G_t)$ \Comment{see Algorithm \ref{algo:graph_building}}
\State $G_s \gets \emptyset$ \Comment{initialize an empty graph for current session}
\For{$i \in \llbracket 1, |s| \rrbracket $} \Comment{for each query in the session}
    \State $G_s \gets BuildLogGraph(i, G_s)$ \Comment{update current session graph with query $i$}
    \State $Gs_t \gets$ {\tt Merge($G_t$, $G_s$, $\alpha$)} \Comment{see Algorithm \ref{algo:merge_graph}}
    \State $belief \gets PageRank( Gs_t )$
    \State \textbf{Yield} $SubjectiveInterestingness(s[i], belief)$ \Comment{as described by Eq.  \ref{eq:si_sum}}
\EndFor
\end{algorithmic}
\end{algorithm}

\section{A first subjective interestingness measure for BI exploration}
\label{sec:subj_interest_dopan}
In this section we describe how a subjective interestingness measure can be defined 
for interactive OLAP explorations. 

\subsection{Definition of the measure}
To construct the Subjective Interestingness measure (denoted by $SI$ hereafter), we follow the same general principle established by De Bie \cite{DBLP:conf/ida/Bie13} and the method presented in Section \ref{sec:DeBie}. In this framework, $SI$ is the ratio of the surprise related to the observation of a pattern and the complexity to understand the pattern in the user point of view, recalled by the following general formulation of $SI$, for a user $u$ seeing a pattern $Q_t$:

\begin{equation}\label{eq:si}
    SI_u(Q_t) = \frac{-log(belief_u^t(Q_t))}{complexity(Q_t)}
\end{equation}


A query being composed of multiple query parts, 
computing the subjective interestingness of a query can
be done using the product of the individual query parts probabilities that compose it. We can therefore rewrite the equation above as:

\begin{equation}\label{eq:si_sum}
    SI_u(Q_t)=\frac{-\sum_{p \in Q_t}\log(belief_u^t(p))}{complexity(Q_t)}
\end{equation}

In the equation above, $Q_t$ is the  $t^{th}$ query of the exploration. $belief_u^t$ if a function dependent of the position $t$ for a specific user $u$. Note that we  assume  the probabilities of query parts to be independent after convergence of the stationary distribution to simplify computations and allow summation of the information content as described in Equation \ref{eq:si_sum}. This assumption is reasonable since the PageRank vector is computed as a probability to reach a given query part after an infinite number of random walks in the graph. After convergence to the stationary distribution, the probability of going to another query part is independent of the previous query part, by definition of the PageRank vector. The probability of a particular sequence of query parts is independent of their order. So, the probability of the sequence of query parts constituting the query is the product of the individual probabilities of the query parts.


\subsection{Query complexity}
Computing interestingness demands a measure that conveys the complexity of the queries. In De Bie's framework, this measure  reflects the difficulty of understanding the pattern. In our case, this "pattern" is actually the query. Previous work already explored such a metric, for instance  for the SQLShare workload  \cite{Jain:2016:SRM:2882903.2882957}. In this paper, the authors use two complexity indicators separately to compare SQL queries originating from different datasets. Those two indicators are discriminant and allow to exhibit different behaviors when applied to the SQL workloads.
Those two indicators are the number of distinct physical operators in the execution plan of a query in the one hand and the query length on the other hand.
As our work is done in the context of BI explorations, we consider in this work multidimensional queries, that may not be phrased in SQL. Indeed, in our experiment we work with queries phrased in MDX.
We therefore decided to use as complexity measure the
number of query parts, since it is correlated to query length (which we have tested for the datasets we used),
and considering that,  even if {\it distinct operators} would be a finer measurement, it is not aligned with the spirit of this complexity measure that should be the complexity as  perceived by the user.

\section{Evaluation of the belief distribution}
\label{sec:tests}

\begin{figure}
    \centering
    \includegraphics[width=170pt]{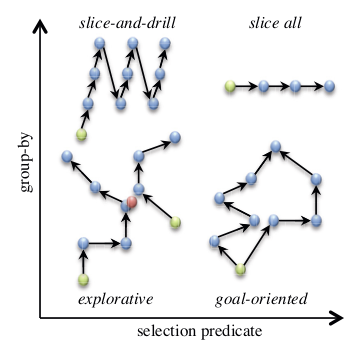}
    \caption{Exploration templates in CubeLoad (from \cite{DBLP:conf/caise/RizziG14}): , {\it ``seed queries in green, surprising queries in red''}.}
    \label{fig:template}
\end{figure}

Our first experiments aim at showing that the belief probability distribution learned with our approach is coherent with what could be expected in realistic exploration situations. To settle such experiments, we envision two distinct use cases. First, we consider 
an ideal environment where all explorations are already categorized into several prototypical user profiles that could be used to bias our model. To this aim, we use the CubeLoad generator \cite{DBLP:conf/caise/RizziG14} which  4  exploratory templates,  illustrated in Figure \ref{fig:template},
 will serve as user profiles.

Second, we use real explorations over an open dataset, called the DOPAN  workload from now on,
where users investigates about energy vulnerability in French R\'{e}gion Centre Val de Loire. We expect these explorations to be more complex and potentially noisy because of queries that are more or less related to the task at hand. This dataset was used in our former work \cite{DjedainiDLMPV19},
where an expert classified  each users' exploration based on their skill expertise.

For each use case, several simulations are conducted to assess that our learned probability distributions behave differently and accordingly to what was expected. 
This section first introduces the experimental protocol for each use case in Section \ref{sec:xp_protocol}, proposes some hypothesis about the expected results in Section \ref{sec:hypothesis} and finally describes and analyses the results in Section \ref{sec:results}.

\subsection{Experimental protocol}\label{sec:xp_protocol}

\paragraph{Evaluation of quality} We will establish our results around two distinct evaluation methods:
\begin{enumerate}
    \item we run a quantitative evaluation that relies on a distance between two probability distributions: the goal is to estimate to which extent they are close and behave similarly. A classical choice could have been to use a Kullback-Leibler divergence, but here we prefer to use the discrete Hellinger distance that has the advantage of being symmetric and bounded in the interval $[0,1]$. The discrete Hellinger distance $H(P,Q)$ compares two discrete probability distributions $P = (p_1, \ldots, p_k)$ and $Q = (q_1, \ldots, q_k)$ as follows:
\begin{equation}\label{eq:hell}
    H(P,Q) = \frac{1}{\sqrt{2}}\sqrt{\sum_{i=1}^k (\sqrt{p_i} - \sqrt{q_i})^2}
\end{equation}
    
    \item we run a qualitative evaluation based on a comparison of plots of average probability distributions presented in  decreasing order. Here, we do not look at a direct comparison of estimated probabilities for a given query part, but we are rather interested in the overall shape of the belief distribution and noticeably how the probability decreases and the long tail behavior.
\end{enumerate}

\paragraph{Implementation} Our approach is implemented in Java using {\tt jaxen} to read 
cube schemas and {\tt Nd4j}\footnote{\url{https://deeplearning4j.org/docs/latest/nd4j-overview}} for simple and efficient matrix computation. The code is open source and available in a {\tt GitHub} public repository\footnote{\url{https://github.com/AlexChanson/IM-OLAP-Sessions}}.

\subsubsection{CubeLoad use case}
We generated a series of 50 explorations using the Cubeload generator 
over the schema of a cube constructed using the SSB benchmark \cite{DBLP:conf/tpctc/ONeilOCR09},
that we split in 2 groups: the first 43 explorations are used to construct the topology graph, and the next 7 are taken from a single CubeLoad template, and are used to define the user profile. 
We run 50 randomized samples to achieve a traditional cross-validation protocol.

\subsubsection{DOPAN use case} The DOPAN workload contains $32$ explorations in total, that have been authored by $10$ master students of a Business Intelligence program. DOPAN contains 3 data cubes with, respectively, 19, 14, and 27 dimensions, and 32, 20, and 58 measures. We expect this dataset to be more challenging than CubeLoad since real users are likely to be less predictable than simulated ones with potential erroneous queries during the exploration. Interestingly, users' explorations are categorised according to the skill expertise in $3$ groups: $B$ users are the less experienced, $C$ users show average skills while $D$ users are supposed to write the most appropriate queries. Noticeably, this dataset's noise and longer explorations can be explained by the behavior of OLAP tools, like Saiku, as they log a new query for each user action (including intermediate drag-and-drops).



\subsection{Hypothesis} \label{sec:hypothesis}

\subsubsection{CubeLoad use case} We expect the 4 templates included in CubeLoad to behave differently.
The {\it slice all} template is a local user model that only explores a small fraction of the data space. 
It is thus expected that when comparing to a baseline distribution probability 
agnostic of any user specific graph, it will maximize this distance. In this case, only a few query parts concentrate most of the interactions with a higher probability, as all queries of the exploration share the same group by set and measure. 
Similarly, as the {\it slice all} template chooses one level in one hierarchy and then only varies the selection predicate, it is expected to show a larger standard deviation than the other templates from one exploration to the next.


On the contrary, the {\it explorative} template simulates a broader exploration of the data space. This template should lead to minimizing its distance with a topology based distribution. In this case, it is expected that there are fewer very improbable query parts but that there are more higher probabilities on most query parts, because of the coverage of the data space by the template.

The {\it goal-oriented} and {\it slice-and-drill} templates are expected to be intermediate states between the two previous templates. Indeed, both models explore the data space more than {\it slice-all}, but are a bit more constrained than {\it explorative}.



\subsubsection{DOPAN use case}

The DOPAN use case is more complex since it deals with real explorations for which we do not know the profile of the users, contrary to CubeLoad. We expect these experiments to confirm the tendencies observed in the CubeLoad dataset, with an ability of our belief distribution to capture the knowledge of the users. However, we expect these results to be less contrasted than what is observed for CubeLoad for two reasons: ($1$) the cubes in DOPAN are more complex  in terms of schema ($2$) the users in the experiments were all trained in the same master program and thus should exhibit some common behviours
, which may not help distinguishing from one profile to the next.

\subsection{Results}\label{sec:results}
\subsubsection{CubeLoad use case}
Table \ref{tab:hellinger} represents the distance between:
\begin{itemize}
\item the PR vector computed over the topology graph $G_t$ (see Section \ref{sec:IM}), 
\item and the PR vector computed over the aggregated graph as produced by Algorithm \ref{algo:merge_graph}, that merges the topology graph $G_t$ with the user specific graph $G_u$. 
In order to bias our model, we gradually modify the parameter $\alpha \in [0,1]$ of the {\tt merge} function in Algorithm \ref{algo:merge_graph} to give more importance to $G_u$ compared to $G_t$.
\end{itemize}

\begin{table}[ht]
\begin{center}
{\scriptsize{
    \begin{tabular}{|l|c|c|c|c|c|c|c|c|c|}
  \hline
  User/$\alpha$ & 0.1 & 0.2 & 0.3 & 0.4 & 0.5 & 0.6 & 0.7 & 0.8 & 0.9 \\
  \hline
Explorative & 0.021 & 0.042 & 0.063 & 0.084 & 0.106 & 0.130 & 0.155 & 0.183 & 0.215\\
Goal Oriented & 0.015 & 0.031 & 0.047 & 0.063 & 0.081 & 0.101 & 0.123 & 0.150 & 0.182\\
Slice All & {\bf 0.073} & {\bf 0.127} & {\bf 0.170} & {\bf 0.209} & {\bf 0.244} & {\bf 0.279} & {\bf 0.315} & {\bf 0.350} & {\bf 0.392}\\
Slice and Drill & 0.022 & 0.044 & 0.066 & 0.089 & 0.114 & 0.139 & 0.167 & 0.200 & 0.236\\
    \hline
\end{tabular}
}} 
    \caption{Average Hellinger distance between the topology graph PR  and the user-specific graph PR, following the  templates of CubeLoad (std. dev. on the order of $10^{-3}$ no represented here)}
    \label{tab:hellinger}
\end{center}
\end{table}

Table \ref{tab:hellinger} details the
Hellinger distance values computed between the two PR's,
for each of the 4 CubeLoad templates, and values of parameter $\alpha$ ranging from $0.1$ to $0.9$.
The first observation from Table \ref{tab:hellinger} is that Cubeload profiles indeed differ in how close they are to the reference distribution. This means that different user activities can be characterized to correspond to different belief by our approach.
We can also observe that the distance between the resulting distributions is proportional to $\alpha$ as expected. Indeed, if $\alpha$ is very low, the biased distribution is very close to the PR topology distribution. The higher $\alpha$, the more characteristics from the user profile are introduced in the transition matrix. Second, and as expected, we notice that the {\it slice-all} profile bears the larger distance with the topology as it only explores a small portion of the possible space, while the other profiles seem to have a comparable behavior in terms of distance. It can be observed that {\it goal-oriented} profile tends to generate the lowest distances to the PR that represents the topology graph. This can be explained by the fact that {\it goal-oriented} is somehow the less constrained simulated user profile as it mainly performs a random walk in the topology graph to a destination query following the schema. In contrast other profiles and noticeably {\it slice-all} and {\it slice-and-drill} restrict more strongly their explorations of the graph to fixed patterns that may contradict probabilities of transitions as observed in most past usages and schema.

\begin{figure}
    \centering
    \includegraphics[scale=0.055]{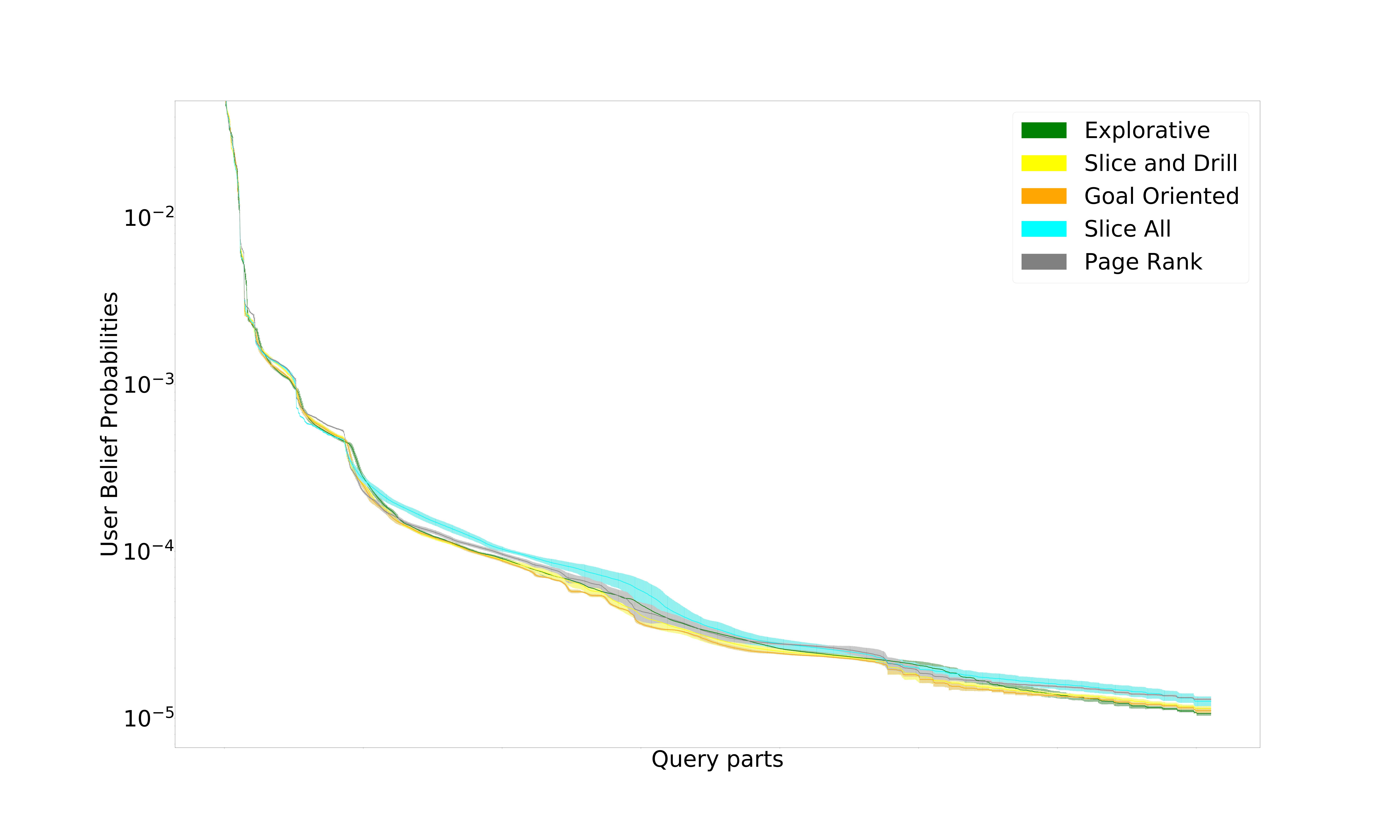}
    \caption{Distribution of probabilities computed by our model for all $4$ user profiles when $\alpha = 0.2$ (log scale). Each plot represents the probabilities of the query parts for one user profile in decreasing order}
    \label{fig:pr_probas02}
\end{figure}
\begin{figure}
    \centering
    \includegraphics[scale=0.055]{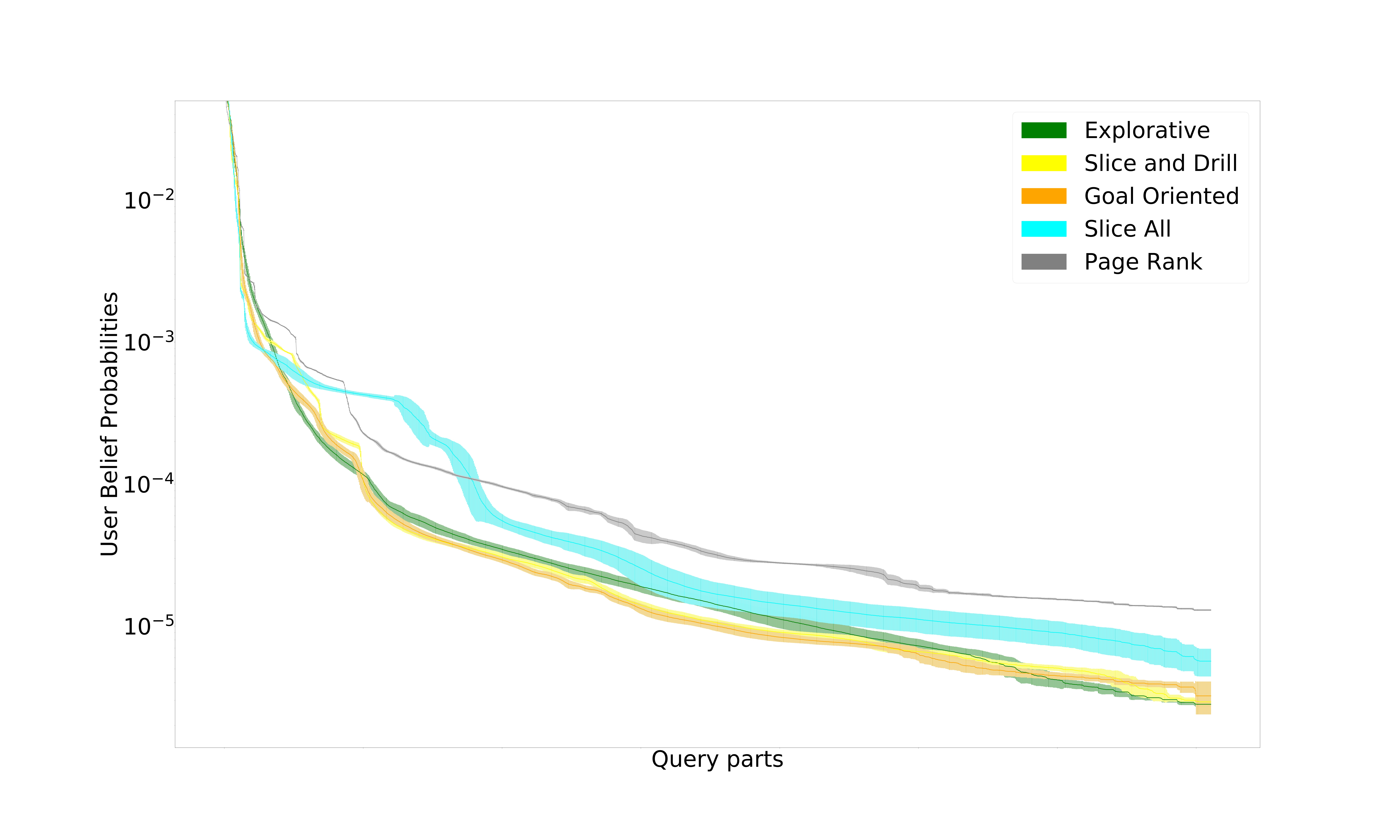}
    \caption{Distribution of probabilities computed by our model for all $4$ user profiles when $\alpha = 0.8$ (log scale). Each plot represents the probabilities of the query parts for one user profile in decreasing order.}
    \label{fig:pr_probas08}
\end{figure}

Figures \ref{fig:pr_probas02} and \ref{fig:pr_probas08} represent, for two distinct values of parameter $\alpha$, the average distribution of probabilities (and their standard deviation) for the $4$ user profiles and the PR distribution corresponding to the topology averaged on 200 tests. As expected, when $\alpha = 0.2$ all distributions heavily tend to mimic the PR distribution. On the contrary, when $\alpha = 0.8$ the difference brought by the user profile become clearly visible. The {\it slice-all} profile tends to have a higher number of higher probabilities and then decreases with successive steps which are characteristic of this profile. Indeed, this profile explores all members at a given depth in a hierarchy and all the corresponding query parts are by construction almost equiprobable: only past usage may modify probability which translates in the observed decrease pattern. Then, as expected the {\it slice-all} profile shows the largest standard deviation in our test, which can be explained by the variability of each exploration from this profile that explore each time a different hierarchy at a different level.\\

Similarly, {\it slice-and-drill} profile exhibits some small but noticeable steps that can be explained by the fact that this profile alternatively navigate between members of a hierarchy at a given level and at some point drill down to query parts from lower level of the hierarchy and that are less likely to be used, as they are more specialized and thus less probable.

Then, it is worth noticing that the PR plot also shows some steps which may traduce the presence of more strongly connected components inside the topology graph where query parts are equiprobable.

Finally, {\it explorative} and {\it goal-oriented} profiles exhibits a more gradual decrease of the probabilities which is again expected as these approaches distributes more evenly their probabilities into more query parts because of their behavior. 



\subsubsection{DOPAN use case}

\begin{table}[ht]
\begin{center}
{\small{
    \begin{tabular}{|l|c|c|}
  \hline
  User/$\alpha$ & 0.2 & 0.8 \\
  \hline
User 03 &0.00134 &0.0586 \\
User 04 &0.00417 &0.135 \\
User 05 &0.000565 &0.00244 \\
User 06 &0.00201 &0.0692 \\
User 07 &0.00560 &0.150 \\
User 09 &0.00423 &0.131 \\
User 10 &0.00367 &0.133 \\
User 12 &0.0000244 &0.00134 \\
User 14 &0.00567 &0.136 \\
User 16 &0.00530 &0.151 \\
  \hline
\end{tabular}
}}
    \caption{Hellinger distance between PR and our biased PR with several user profiles on the DOPAN dataset}
    \label{tab:hellinger_dopan}
\end{center}

\end{table}

Similarly to Table \ref{tab:hellinger}, Table \ref{tab:hellinger_dopan} indicates the distances between the PR topology distribution and distributions obtained by biasing the topology graph with users explorations on the DOPAN cubes. For the sake of simplicity, Table \ref{tab:hellinger_dopan} only presents the Hellinger distances computed for 2 distinct values of parameter $\alpha$ that are either topology oriented ($\alpha = 0.2$) or user oriented ($\alpha = 0.8$).

It can be first noticed that the distance values are much smaller than in the case of CubeLoad explorations as expected. This is explained by the higher complexity of the DOPAN cubes combined with fewer explorations per user, in contrast to the CubeLoad experiments that were conducted with more explorations over a simpler SSB cube.
This indicates that our modeling of belief therefore correctly accounts for the complexity of the query space.

Noticeably, two users exhibit very low distances to the reference belief. By manually reviewing those user logs we observed that they had very short explorations compared to the other users. As a consequence, a small user log will not change much the aggregated graph when mixed with the topology graph and therefore have almost no influence on the PR vector and on the computed distances.
In other words, 
the probability to use a specific query part for a user with little experience is dictated by the schema and general user navigational habits.

\begin{figure}
    \begin{center}
        \includegraphics[scale=0.04]{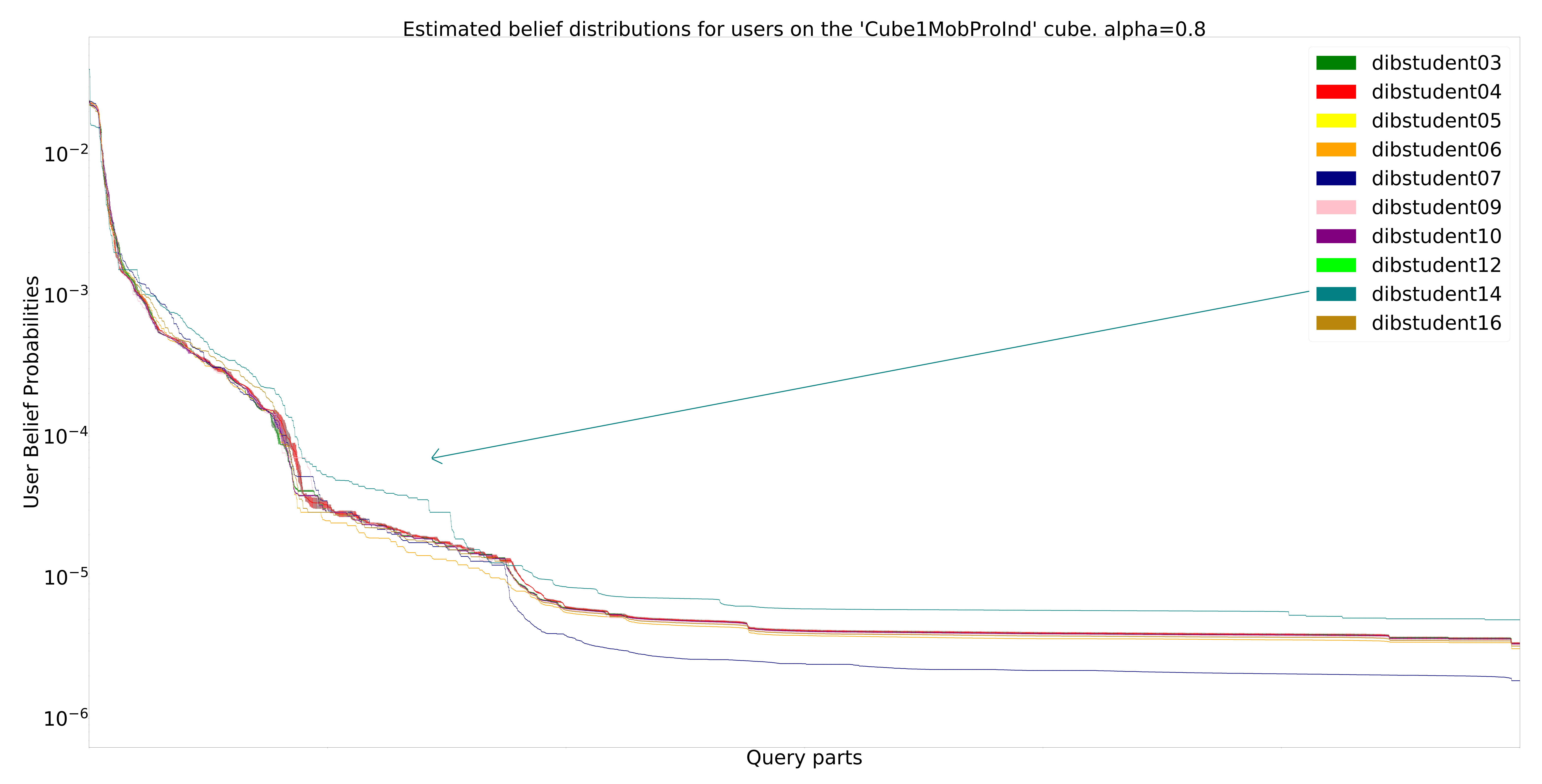}    
        \caption{Distribution of probabilities over the first (and most used) cube of the DOPAN dataset. Query parts are ordered according to their decreasing probability value for each student. \textit{Note that the plot has been limited to the first $3000$ queries for the sake of readability.}}
        \label{fig:belief_dopan}
    \end{center}
\end{figure}

In Figure \ref{fig:belief_dopan} we display in the same way as Figure \ref{fig:pr_probas02} and Figure \ref{fig:pr_probas08} the belief distribution of the user but here on the real explorations over the first cube of the Dopan dataset. Interestingly, we can observe that student $14$ behavior is reminiscent to the behavior exhibited by CubeLoad's \textit{Slice All} profile. By reviewing the user's explorations we found that they change the selection predicate over several queries while keeping measures and group by set elements from previous queries. This behavior is  very similar to the one described by a \textit{Slice All} pattern, and indicates that our modeling of belief 
can help differentiating specific ways of exploring.


\section{Exploration evaluation based on Subjective Interestingness}
\label{sec:test_si_dopan}

We use  Algorithm \ref{algo:eval} to compute the subjective interestingness (SI) on the explorations of our two datasets. 
The aim of these tests is to show how SI
relates to prototypical user behaviors (Cubeload use case) 
and to real user explorations  (DOPAN use case).
We start by describing the protocols for the two use cases and then
comment our experimental results.

\subsection{Experimental protocol}

In both use cases, we use $\alpha=0.9$ for Algorithm \ref{algo:eval}
to compute the SI incrementally for each explorations, to better account
for the user (simulated or real) behavior.
We now explain the difference in protocol for the two uses cases.

\paragraph{CubeLoad} 
Explorations generated with CubeLoad correspond to prototypical
behaviors (profiles) of users navigating a datacube.
For our first experiment, we generate $50$ explorations for each different CubeLoad profile. 
We first plot the accumulated number of unique query parts used at each moment of the exploration to understand how our complexity measure behaves.
We then run  Algorithm \ref{algo:eval} 
on each explorations to compute the subjective interestingness per query. For each  profile, we isolate the current explorations from the others generated. All the other explorations of the same profile will be used as a user past log. The results are finally aggregated per profile and query position in the exploration, to compute the mean and the standard deviation of SI. We display the results in the form of a line plot with error bars representing the standard deviation. The query position on the $x$ axis represents different moments of the exploration. Each line is a cubeload profile representing the mean behavior of all explorations generated using this profile. Our aim is to see how SI behaves along the explorations and allows to characterize the prototypical profiles.

\paragraph{DOPAN} For the DOPAN use case, we have at our disposal $32$ real user  explorations.
We will focus on the 22 explorations over the first cube of DOPAN, since SI cannot be compared across cubes with different schemas.
Contrary to the simple SSB cube, that one has 32 measures and 19 dimensions,  making it a much larger space to explore.
Our protocol is a bit different than the one used for CubeLoad since we have not classified the explorations in how they follow a particular pattern.
However, each exploration has been tagged by professors with labels B, C or D, depending on analyst’s skills. Label D corresponds to good explorations, clearly following an information need, investigating it and containing coherent queries. Students producing such explorations are considered to have analysis skills. Contrarily, label B denotes those of the students that produced poor explorations, with less contributive queries, typically switching topics, with no clear information need. Label C corresponds to students that are learning analysis skills, but still produce middle-quality explorations.

Like for the CubeLoad use case, we first plot the accumulated number of unique query parts used at each moment of the exploration.
We run Algorithm \ref{algo:eval} on each exploration to compute SI per query, using, for the user history to build $G_s$, an exploration consisting of 
the past queries of the exploration and, whenever possible, the queries of other explorations of that user. Note that users have done different number of explorations, each of different sizes, and therefore our protocol faithfully represent genuine user activities.
We use the same line plot as for CubeLoad to show the SI of each query of the explorations, to analyze how SI varies in the exploration and detects instantaneous analysis behavior. 
We then group explorations per skill labels and plot the mean and confidence interval at 95\% of SI for the 3 groups, again with line plot.
We finally compute the rank correlation between SI and the 
skill.



\subsection{Results}

\subsubsection{CubeLoad use case}

\begin{figure}
    \centering
    \includegraphics[scale=0.23]{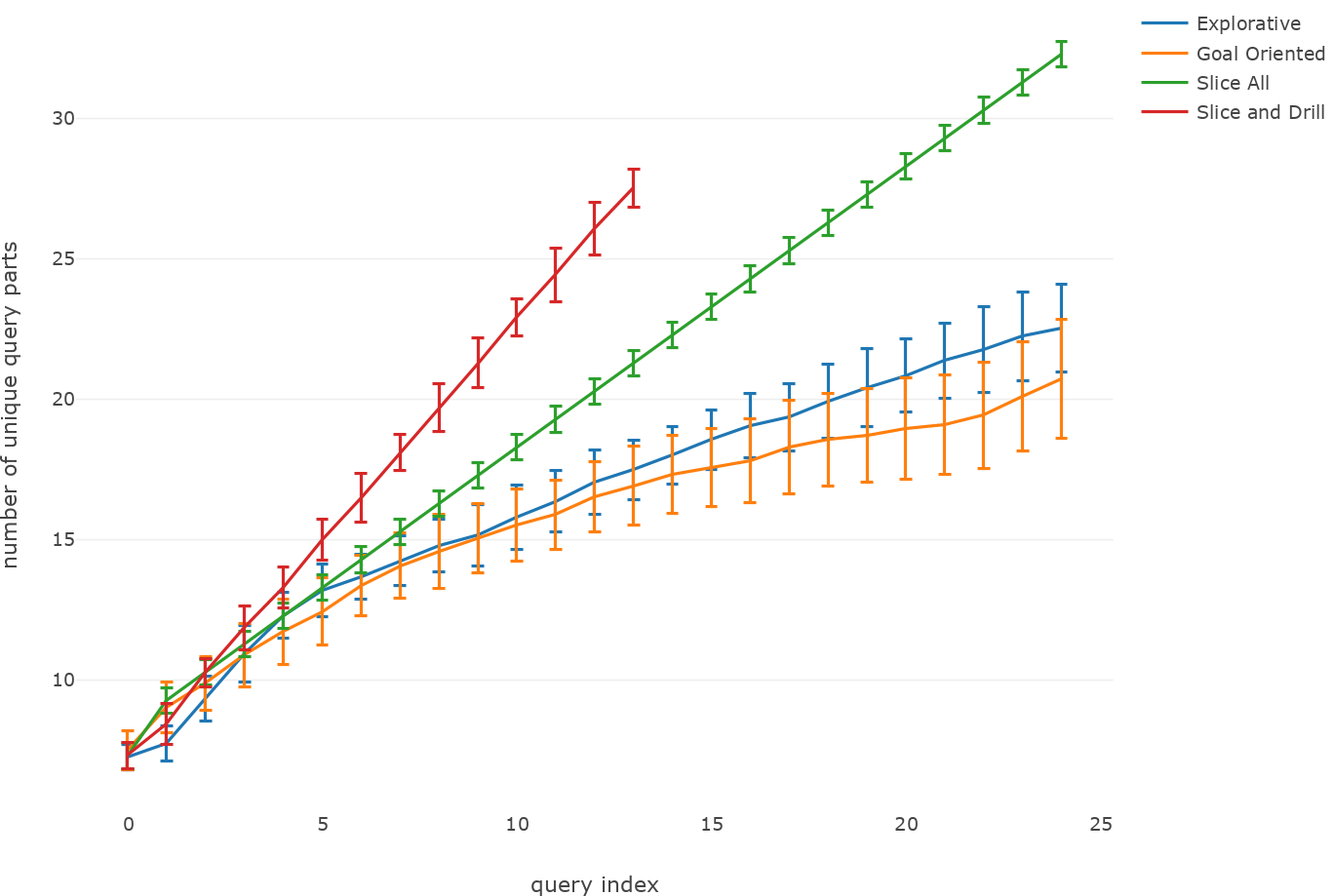}
    \caption{Cumulated number of unique query parts by CubeLoad template for each query index in the explorations}
    \label{fig:cubeload-queryPartNumber}
\end{figure}

Figure \ref{fig:cubeload-queryPartNumber} shows the accumulated number of unique query parts used at each moment of the exploration. This measure increases monotonically since each newly generated query adds one or more parts, some being already seen and counted, allowing to rank the 4 profiles.
As expected, the slice and drill profile generates the more never encountered query parts since this pattern necessarily goes in one new direction, either by drilling or slicing.
It is followed by the slice all pattern, that only add new slices. 
Then comes the two profiles that are constrained by either a goal query (goal oriented) or a "surprising" query (explorative), and as such may not necessarily add never seen query parts since they are forced to stay around those queries. The goal oriented has a bit less new query
parts since it is constrained all along the exploration, while explorative is only constrained half of it.

\begin{figure}[h]
    \centering
    \includegraphics[scale=0.25]{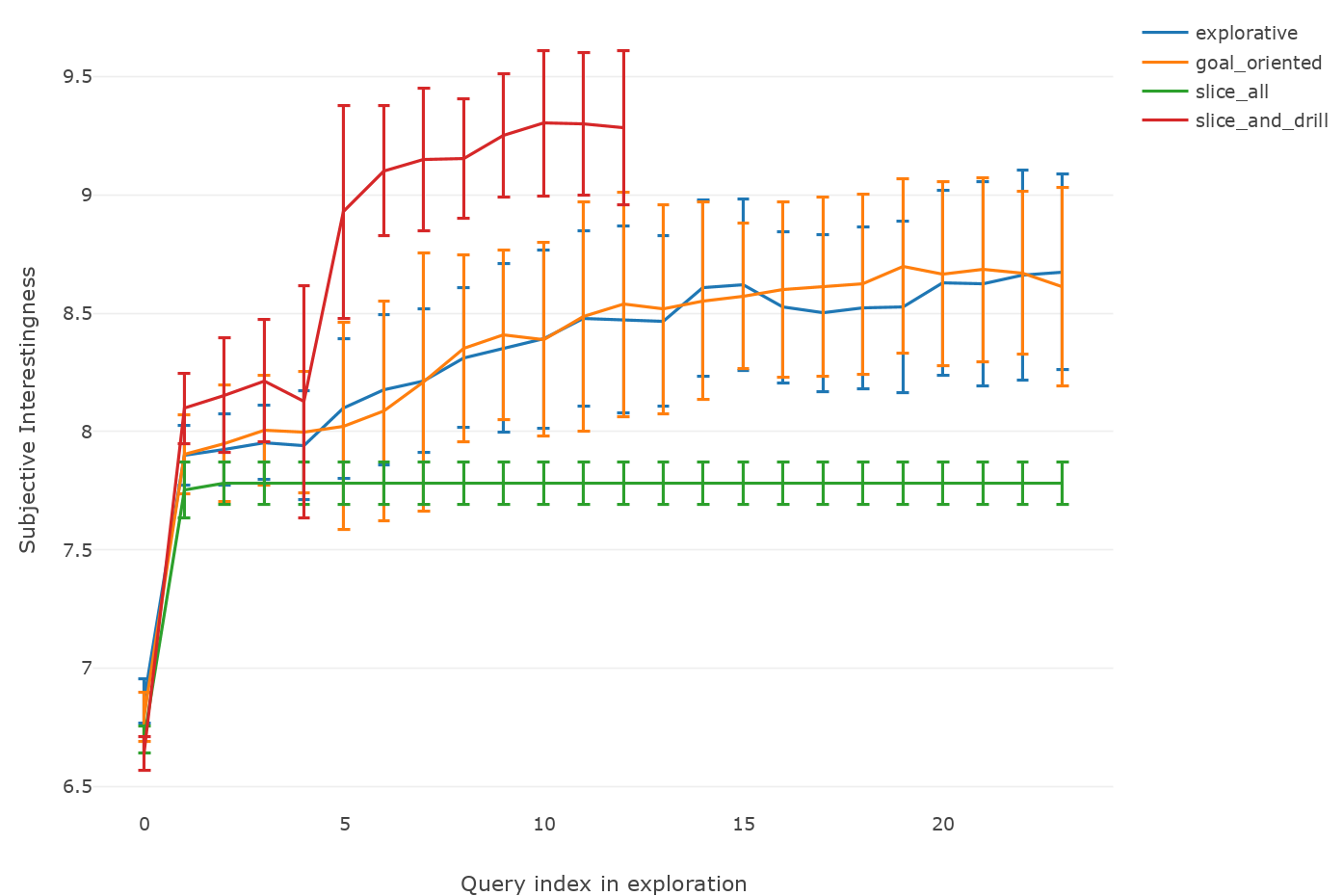}
    \caption{Subjective Interestingness for each cubeload  profile}
    \label{fig:cubeload_SI_profile}
\end{figure}

As can be observed in Figure \ref{fig:cubeload_SI_profile}, SI is capable of discriminating the cubeload profiles, except for the \textit{goal oriented} and \textit{explorative} profiles that cannot be distinguished. This behavior is expected since both profiles 
exhibit changes to selection and group by attribute at each query. Both behaviors  will therefore generate quite large amounts of information and thus will hardly be  distinguished with SI.

The \textit{slice all} profile shows the least amount of Subjective Interestingness because of the way its queries are generated. At each step, the query parts are very minimally altered, thus generating less surprise while keeping a constant 
query complexity. The combination of these two behaviors cause the Subjective Interestingness to be almost constant across all queries of an exploration for this profile.
The \textit{slice and drill} profile is clearly the profile generating the most subjectively interesting queries. 
Compared to the other profiles, slice and drill generates the greater variety of query parts since it systematically moves by using new query parts, without remaining at the same group by level (like \textit{slice all}) and without  being
attracted to some queries (like \textit{explorative} or \textit{goal oriented}). 
Being both constrained, albeit not in the same way, explorative and goal oriented profiles exhibit an intermediate behavior and cannot be easily distinguished with SI.

\subsubsection{DOPAN use case}

\begin{figure}[h]
    \centering
    \includegraphics[scale=0.2]{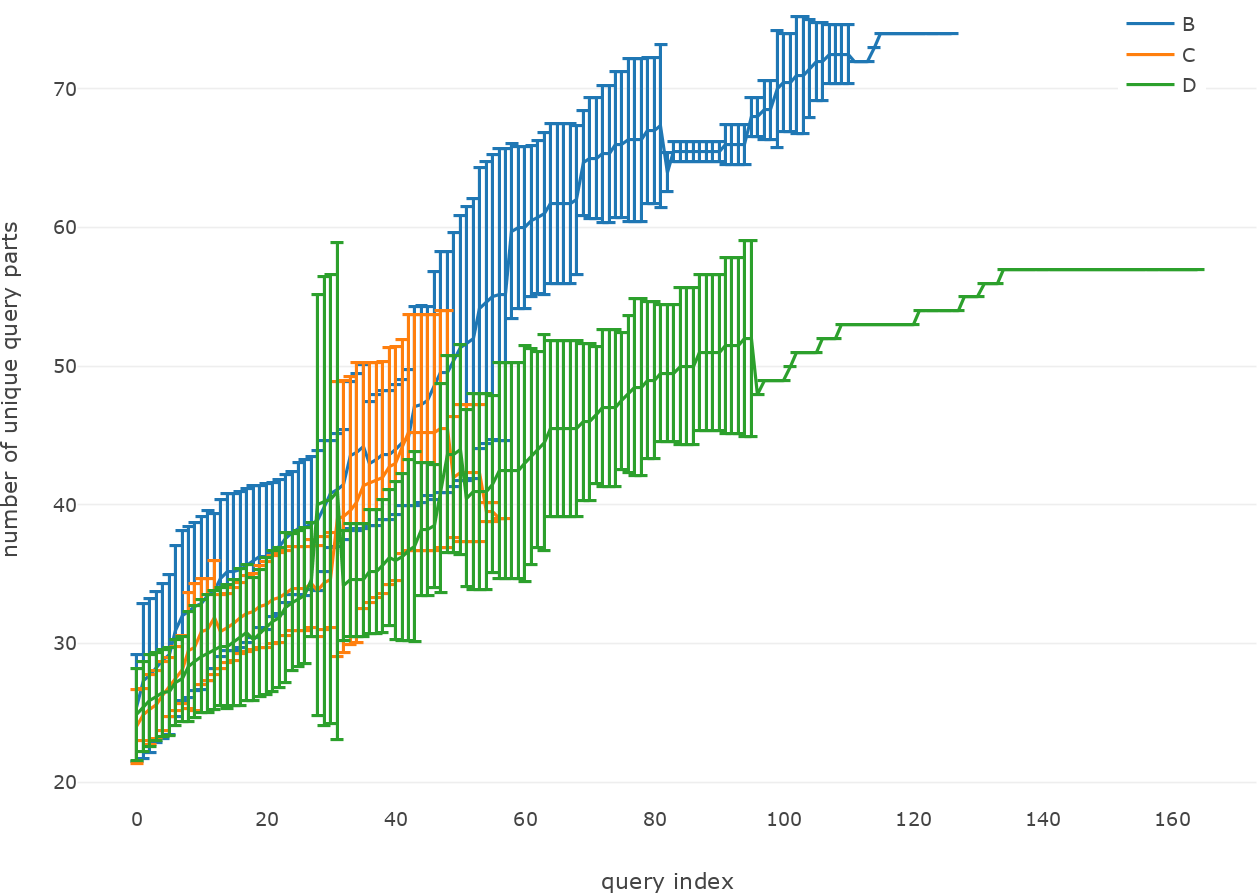}
    \caption{Cumulated number of unique query parts by skill for each query index in the explorations.}
    \label{fig:cumulated_query_parts}
\end{figure}

Figure \ref{fig:cumulated_query_parts} plots the cumulated number of unique query parts by skill (B, C or D) for each query in the explorations. 
Clearly, B labelled explorations have more never seen new query parts, while D labelled explorations have less never seen query parts. The increase rate is also more pronounced for  B explorations, and  higher than for  D  explorations. This is explained by the fact that lower skill explorations exhibit a more explorative and erratic behaviour that periodically tries 
new directions to explore. Manually reviewing the B explorations, we indeed have found that  their behavior is sometimes reminiscent of the slice and drill profile described in Figure \ref{fig:template}.
On the contrary, D labelled explorations tend to produce less query parts, since these explorations exhibit the behavior 
of a user knowing how to formulate efficiently minimal queries to get to their objective. 

Category C explorations have a mixed behavior with a similar behavior as category D at the beginning and then gradually converging to a situation where they produce as many query parts as the beginner explorations. 
Interestingly, category C also produces shorter explorations, as shown in Figure \ref{fig:cumulated_query_parts},
as these explorations were deemed to be executed by a "user acquiring analytically skills" ;  they will choose useful measures, use relevant selection predicates but their explorations might be cut short because they did not manage to answer a business questions or did not go the extra step to understand discrepancies in the data they find (e.g., by drilling down).
Finally, Figure \ref{fig:cumulated_query_parts} reflects with the cumulated number of query parts, the spread of each exploration skill profile in our query parts graph, which  directly impacts our Subjective Interestingness measure.


\begin{figure}[h]
    \centering
    \hspace{-1cm}
    \includegraphics[scale=0.17]{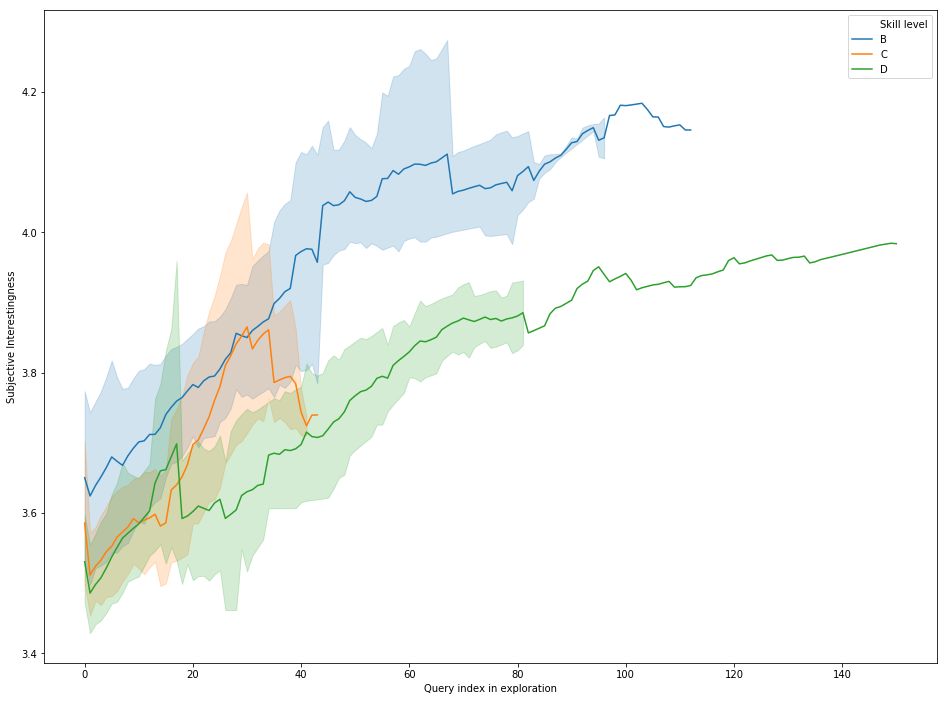}
    \includegraphics[scale=0.13]{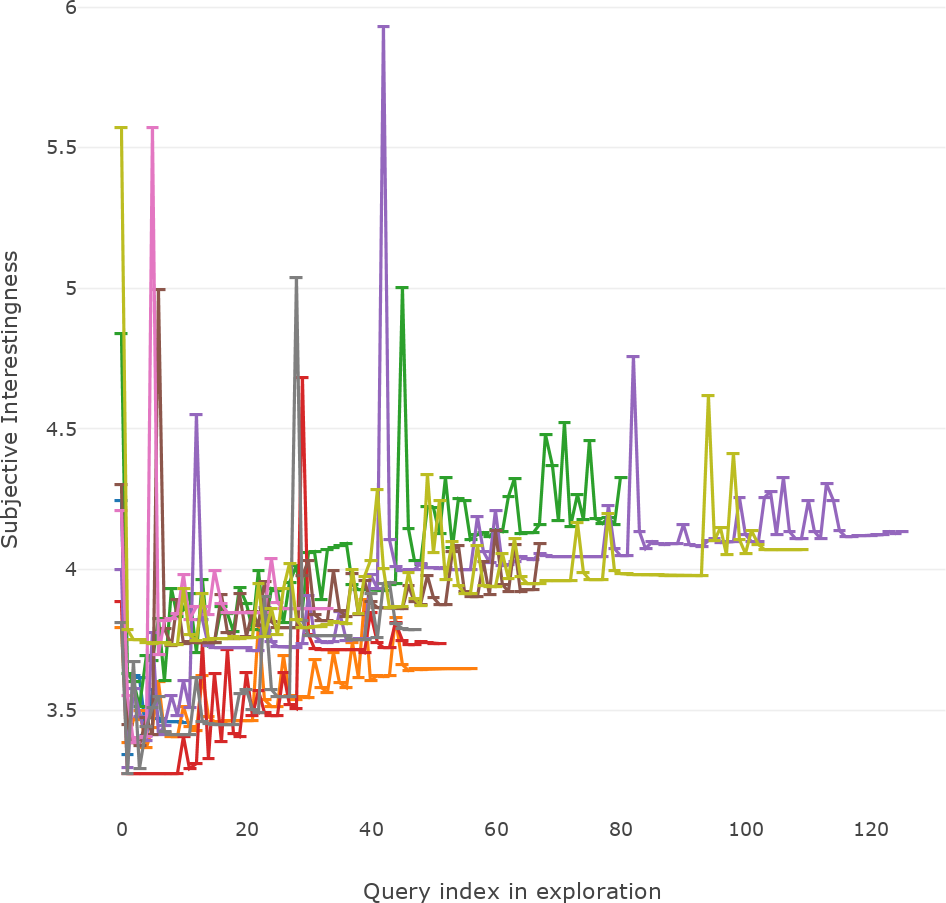}

    \includegraphics[scale=0.15]{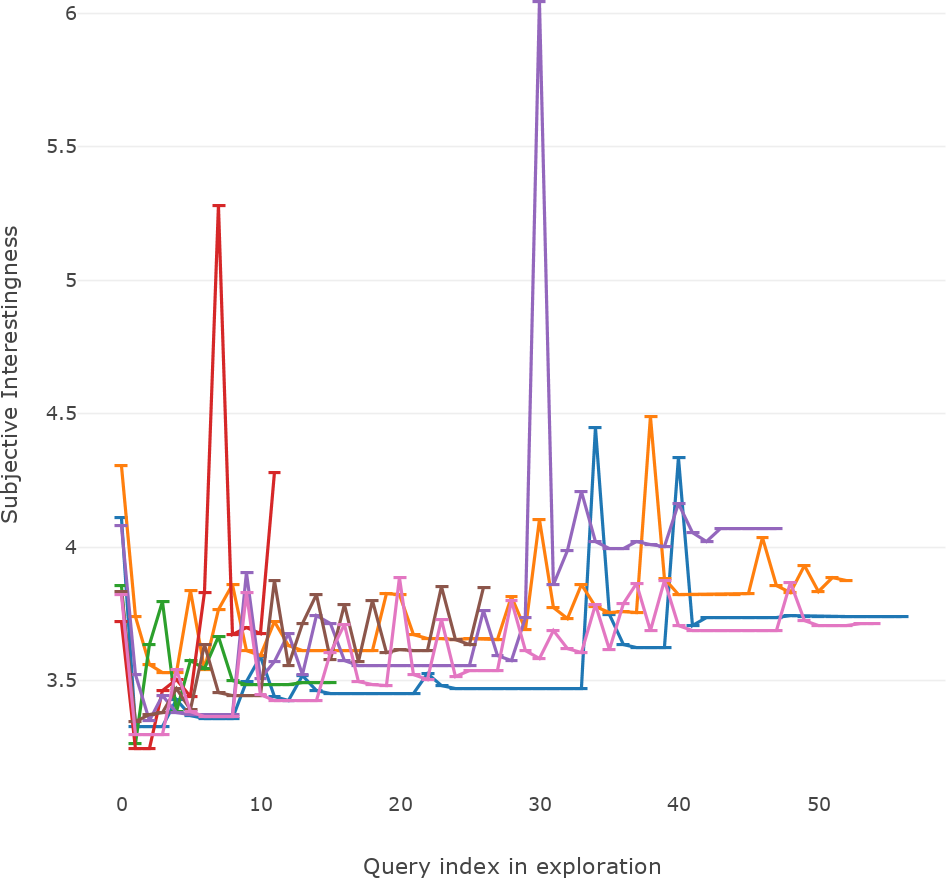}
     \includegraphics[scale=0.15]{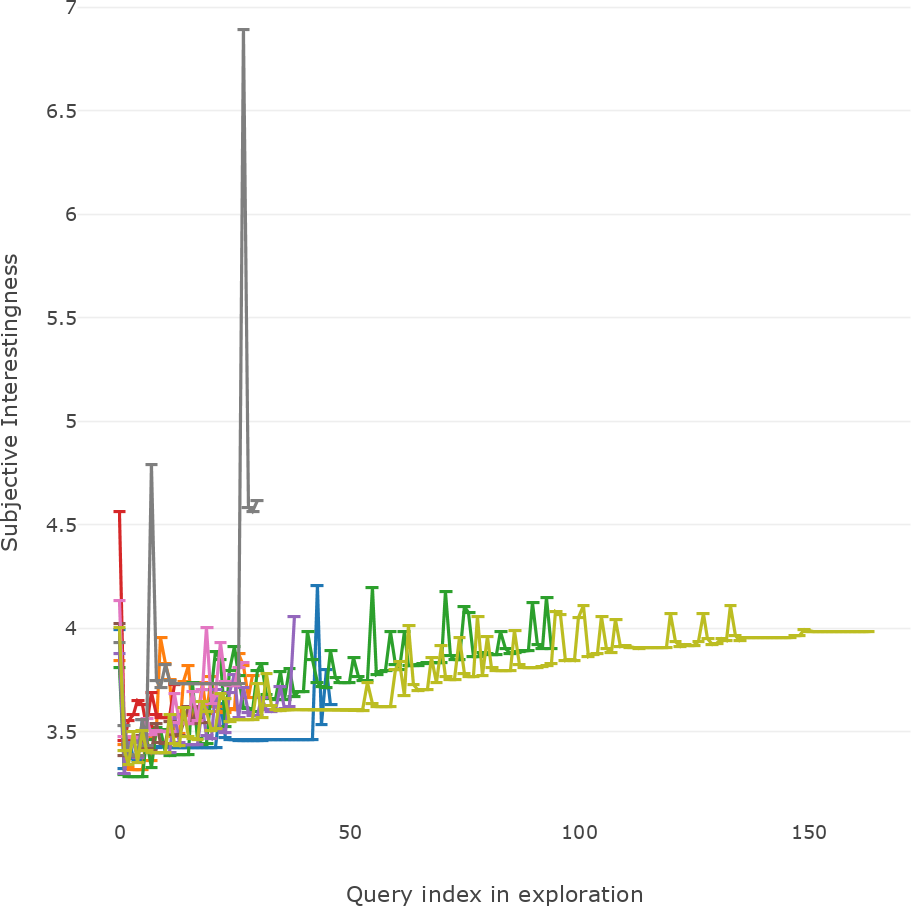}
    \caption{Average and confidence interval of SI by skill for each query index in the explorations (top left), SI for all explorations of skill B (top right), C (bottom left) and D (bottom right), each color being an individual exploration.}
    \label{fig:query_SI_profile}
\end{figure}

Figure  \ref{fig:query_SI_profile} (top left) 
shows  the average SI with  confidence interval of each query of the explorations,  grouped by skill labels and presented by query position in the exploration.
Noticeably, SI characterizes B labelled explorations with the highest average score and D labelled explorations with the lowest, while C labelled explorations are in between.
High SI corroborates the explorative nature of B explorations, where users struggle to find their way in the multidimensional space by selecting unseen query parts somewhat erratically. On the contrary, low SI corroborates the more "focused" nature of D explorations, where users are more pragmatic in their choice of query part to express classical OLAP operations like roll-up or drill-down. Doing such choices, i.e., selecting query parts close to the ones already employed mechanically lowers SI, since the belief attached to those new parts is high and therefore their surprise is low.
Regarding C explorations, they were labelled a such because they exhibit behaviors coming partly from unskilled explorations and partly from skilled ones. While SI is not conceived to discriminate user skills,
it can still position those intermediate behaviors between the two extremes.

Figure  \ref{fig:query_SI_profile} (top right and bottom) shows the SI for the queries of each exploration, separated by skills.
It is immediately noticeable that there are sudden and short spikes in the SI. By looking at the specific queries where those spikes occur, 
we could see that the users suddenly add large numbers of selection predicates. By design, our measure can
detect such behaviors since adding a high number of parts, even if they have a high probability of being selected, result in high SI due to how probabilities are used to compute surprise (see equation \ref{eq:si_sum}) and because of the sudden increase in the complexity of the query. 
The queries obtained with such a burst of new query parts are likely to be informative and correctly detected by our measure.


%

Finally, we have validated these observations with  a rank correlation test between skills (B, C, D) and the Subjective Interestingness measure that outputs a score of $-0.28$. This result confirms that there is a correlation between the skill category and our Subjective Interestingness Measure: the lower the category, the higher the Subjective Interestingness.






\section{Related Work} \label{sec:related}

Our work deals with subjective interestingness and how to define such a measure by learning a belief distribution from users' past activities in the context of Business Intelligence. This section  presents some interestingness measures, and how they have been used in the context of recommendation and exploratory data mining.

Defining a good Interestingness measure has kept interested researchers for a long time in the context of data mining. Indeed, there exists numerous tasks, for example in pattern mining, for which it is critical to be able to filter out uninteresting patterns such as item sets or redundant rules, to control the complexity of the mining approaches and increase their usability. 

In \cite{Brijs2004DefiningIF, DBLP:journals/csur/GengH06}, the authors identify two main types of interestingness measures. Objective measures are based only on data, which corresponds to quality metrics such as generality, reliability, peculiarity, diversity and conciseness. For instance, directly measurable evaluation metrics such as support confidence, lift or chi-squared measures in the case of  association rules \cite{BostonCollege-Alvarez-2003}. 

On the contrary, subjective measures consider both the data and the user and characterize the patterns' surprise and novelty when compared to previous user knowledge or expected data distribution. The first work on the topic of subjective interestingness is certainly \cite{DBLP:conf/kdd/SilberschatzT95} that is restricted to the pattern mining domain. In \cite{DBLP:conf/ida/Bie13, IDEA:2018}, the author extends this notion to any explorative data mining task and represents interestingness as a ratio between information content and complexity of a discovered pattern being it an itemset, a cluster or a query evaluation result (see Section \ref{sec:DeBie} for more formal details). In \cite{IDEA:2018}, De Bie defines the subjective interestingness as a situation where a \begin{quote}
    ``user states expectations or beliefs formalized as a ‘background distribution’. Any ‘pattern’ that contrasts with this and is easy to describe is subjectively interesting''.
\end{quote}

The authors in \cite{DBLP:journals/csur/GengH06} consider also semantic measures of interestingness, based on the semantics
and explanations of the patterns like utility and actionability. This latter property of actionability is not meaningful in our case where, as stated by De Bie \cite{IDEA:2018}, we consider situations \begin{quote}
``where the user is interested in exploring without a clear anticipation of what to expect or what to do with the patterns found''.
\end{quote}

De Bie's framework \cite{DBLP:conf/ida/Bie13} is usually used to model user belief about some \textit{data}. The hypothesis is that the user has beliefs about all of the data and is interested by anything that is surprising according to her beliefs. This will usually be a piece of data whose properties greatly contradict the user's prior belief. This work is very similar to what Sarawagi did for multidimensional data exploration \cite{DBLP:journals/vldb/Sarawagi01}. In her work, the user's previous observations about parts of the data is used to estimate the most probable cube instance in the user's brain. A dissimilarity to the actual cube instance data is then computed in order to recommend surprising subsets. Both approaches compute a kind of information gain conditioned by the knowledge of what has been seen by the user while he explored the data. This gives the system the ability to suggest the best action that will provide the most information out of the data to the user.

Modelling the probability over the user \textit{intents} would be another subjective approach. This is well illustrated by the Bayesian Information Gain method, as shown in various works by Wanyu \cite{DBLP:conf/chi/LiuDBR17} for instance. With this approach, the system makes hypothesis about the user's goal and test them by subjecting her to experiments. The goal is to find the experiment which would yield most information about the user's goal to the system, to help her reach it faster. Both approach seems quite complementary as De Bie's approach makes possible query recommendation through \textit{interesting data discovery} and Wanyu's approach instead allows similar recommendation through \textit{intent discovery}. \nico{Interestingly, these intents are probability distribution over elements of knowledge, while other works have focused on capturing long or short term intents related to topics for data exploration such as \cite{DBLP:conf/caise/DrushkuALMPD17}.}

Recently, in \cite{DBLP:conf/icde/PuolamakiOKLB18} the authors propose a data exploration study based on De Bie's FORSIED framework \cite{foresied, IDEA:2018} that pairs a high level conjunctive query language to identify groups of data instances and expresses belief on some real-valued target attributes, based on location and spread patterns. This work is close to our proposal but expresses belief
on a summary of the data.

In general, most of recent work on De Bie's framework \cite{DBLP:conf/icde/LijffijtKDPOB18, DBLP:journals/ml/LeeuwenBSM16} instantiate his framework to discover subjectively interesting pattern for different kind of data spaces. As a consequence, papers detail new pattern syntax as well as other statistics computed on the pattern extension. For example, in \cite{DBLP:journals/ml/LeeuwenBSM16}, De Bie's uses his framework to find subjectively interesting dense graph patterns: each pattern is a set of nodes and the statistic is the average degree of the edges in the pattern's nodes. In \cite{DBLP:conf/icde/LijffijtKDPOB18}, the pattern used are subgroups of instances described by conjunctions of categorical descriptive attributes, while the statistics are the mean and the covariance of the subgroup for any number of real-valued target attributes.\\


In the context of data cube exploration, to the best of our knowledge there is no final and consensual  interestingness measure or belief distribution elicitation method, while there exists  measures that are closely related. Measures have been defined as unexpectedness of skewness in navigation rules and navigation paths \cite{KUMAR2008} and computed as a peculiarity measure of asymmetry in data distribution \cite{KLEMETTINEN1999}. In \cite{DBLP:conf/sbbd/FabrisF01}, the authors define interestingness measures in a data cube as a difference between expected and observed probability for each attribute-value pair and the the degree of correlation among two attributes. 
In \cite{DBLP:conf/vldb/Sarawagi00}, Sarawagi describes a method that profiles the exploration of a user, uses the Maximum Entropy principle and the Kullback-Leibler divergence as a subjective interestingness measure
to recommend which unvisited parts of the cube can be the most surprising in a subsequent query.

In \cite{DjedainiDLMPV19, DBLP:conf/adbis/DjedainiLMP17} the authors use supervised classification techniques to learn two interestingeness measures for OLAP queries: ($1$) focus, that indicates to what extent a query is well detailed and connected to other queries in the current exploration and ($2$) contribution that indicates to what extent a query contributes to the interest and quality of the exploration.

Finally, interestingness and related principles have been studied in the context of recommendation but more widely used for evaluation rather than the recommendation itself \cite{DBLP:journals/tiis/KaminskasB17}. Interestingness is reflected based on $4$ main criteria such as diversity, serendipity, novelty, and coverage, in addition to traditional accuracy measures. 

In the context of OLAP query recommendation, several recommendation algorithms have been proposed that take into account the past history of queries of a user either based on a Markov model \cite{DBLP:conf/dawak/Sapia00} or on extracted patterns \cite{DBLP:journals/dss/AligonGGMR15}. Noticeably, \cite{DBLP:journals/dss/AligonGGMR15} quantifies how distant is the recommendation from the current point of exploration to evaluate the interestingness of each candidate query recommendation.



\section{Conclusion} \label{sec:conclusion}

This paper addresses the 
question of determining what is interesting for a specific user during an interactive  
exploration of a multidimensional cube. To that extent, the paper draws a 
parallel with De Bie's Forsied framework \cite{DBLP:conf/ida/Bie13}, and defines a subjective interestingness measure for a query as a ratio between the surprise expressed 
through this query and its complexity.
%
Defining such a measure raises $3$ main challenges: ($1$) how to model and learn the prior user belief as a probability distribution 
to model surprise?
($2$) How to efficiently recompute this prior belief after each user query? And, ($3$) how to implement a realistic Subjective Interestingness measure that captures the complexity of each query?

Our measure definition takes advantage of 
the specificities  of Business Intelligence explorations of multidimensional data cubes.
We represent the prior knowledge of a specific user as a directed graph of query parts that relies: on former users' explorations, as a proxy of what the current user might find interesting, on the cube schema, that indicates how prior knowledge is structured, and finally on this specific users' past activity. 
Finally, the user belief is derived from this query parts graph as the stationary probability distribution of a PageRank algorithm.
Experiments conducted on simulated realistic user explorations 
or on real user explorations 
show that the observed belief distribution and subjective interestingness values are aligned with prior knowledge on these datasets.
This first work on the definition of a subjective interestingness measure in Business Intelligence shows that query parts offer a reasonable proxy
to learn an appropriate model of user belief and determine what is interesting in an exploration.

However, in De Bie's framework, the belief is expressed on the extension of the data and not on the intention of the way of characterizing the data subgroup. A first major extension to our work will consist in proposing a belief and subjective interestingness measure on the value of each cube cells. 
Second, we plan to investigate how to define such an interestingness measure for less structured databases, like relational (non multidimensional) databases, or data lakes.
%
Finally, we aim at studying how user belief modeling and subjective interestingness measures combine with
higher level intentional languages to query and learn from the data \cite{DBLP:conf/dolap/VassiliadisM18}.

\bibliographystyle{ACM-Reference-Format}
\bibliography{krista_biblio}

\end{document}